\documentclass{article}
\usepackage{spconf,amsmath,graphicx}
\usepackage{color}
\usepackage{enumitem}
\usepackage{multirow}
\usepackage{etoolbox,siunitx}
\robustify\bfseries
\usepackage{booktabs}
\usepackage{balance}
\usepackage{amssymb}
\usepackage{bm}
\usepackage{arydshln}
\usepackage{caption}
\usepackage{cite}
\usepackage{hyperref}

\usepackage{amssymb}%
\usepackage{pifont}%

\def\CC{{\mathbb C}}
\def\RR{{\mathbb R}}

\usepackage{caption}
\let\OLDthebibliography\thebibliography
\renewcommand\thebibliography[1]{
  \OLDthebibliography{#1}
  \setlength{\parskip}{0.5pt}
  \setlength{\itemsep}{0.3pt plus 0.1ex}
}
\title{Neural Speech Enhancement with Very Low Algorithmic Latency and Complexity via Integrated Full- and Sub-Band Modeling}
\name{Zhong-Qiu Wang$^1$, Samuele Cornell$^2$, Shukjae Choi$^3$, Younglo Lee$^3$, Byeong-Yeol Kim$^3$, Shinji Watanabe$^1$}
\address{
$^1$Language Technologies Institute, Carnegie Mellon University, Pittsburgh, USA \\
$^2$Università Politecnica delle Marche, %
Italy \,\,\,\, $^3$Hyundai Motor Group and 42dot Inc., Seoul, Korea \\
{\small\texttt{wang.zhongqiu41@gmail.com}}\vspace{-0.6cm}
}
\vspace{-0.3cm}

\begin{document}
\ninept

\maketitle

\begin{abstract}
We propose FSB-LSTM, a novel long short-term memory (LSTM) based architecture that integrates full- and sub-band (FSB) modeling, for single- and multi-channel speech enhancement in the short-time Fourier transform (STFT) domain.
The model maintains an \textit{information highway} to flow an over-complete input representation through multiple FSB-LSTM modules.
Each FSB-LSTM module consists of a full-band block to model spectro-temporal patterns at all frequencies and a sub-band block to model patterns within each sub-band, where each of the two blocks takes a down-sampled representation as input and returns an up-sampled discriminative representation to be added to the block input via a residual connection.
The model is designed to have a low algorithmic complexity, a small run-time buffer and a very low algorithmic latency, at the same time producing a strong enhancement performance on a noisy-reverberant speech enhancement task even if the hop size is as low as $2$ ms.
\end{abstract}

\vspace{-0.2cm}
\section{Introduction}
\vspace{-0.2cm}

Deep learning has dramatically advanced speech enhancement in the past decade~\cite{WDLreview}.
However, current enhancement models reporting strong performance usually consist of many layers of convolutional, recurrent or self-attention blocks.
They are often computationally-intensive, resource-demanding and suffer from large processing latency not suitable for online real-time enhancement, with low-latency, low-complexity enhancement largely being under-explored.
These issues prevent the deployment of modern neural speech enhancement models into real-world products such as hearing aids which usually have very limited computing capabilities, and dramatically limit the potential application range of deep neural network (DNN) based enhancement.
As is suggested in~\cite{Tzinis2020, Wang2022VeryLowAlgorithmicLatency}, an ideal neural speech enhancement system needs to have a small model size and consume a small amount of memory, computation and energy at training and inference time, meanwhile achieving strong enhancement performance with very low processing latency\footnote{Processing latency consists of algorithmic latency resulting from algorithmic design (e.g., the use of overlap-add) and hardware latency for the computation at each frame~\cite{Wang2022VeryLowAlgorithmicLatency}.}.

Many recent neural speech enhancement studies~\cite{Luo2019, Liu2020, Luo2020, Pandey2021, Yang2022TFPSNet, Wang2022GridNet} have a particular focus on using a smaller model size to achieve stronger enhancement performance. %
Although very small model sizes are certainly desirable, in most modern edge devices a model size below $20$ megabytes (MB) is typically satisfactory as the storage and RAM are usually much larger.
The more pressing issues, we believe, are in the run-time memory cost, algorithmic complexity, and computation requirements when performing one-frame-in, one-frame-out enhancement in a real-time fashion.
Solving these issues requires major changes to many current DNN architectures.
For example,
\begin{itemize}[leftmargin=*,noitemsep,topsep=0pt]
\item Attention mechanism~\cite{Vaswani2017,Subakan2021, Subakan2022aRESepFormer, Zhang2022Axial, Pandey2022}, especially in its original form which attends to past frames to capture long-range context, is not ideal for low-complexity, online enhancement, since it needs to buffer many past frames and hence has a sizable memory cost;
\item Although two-dimensional (2D) convolution (Conv2D) features a small number of parameters and has been popular in UNet-based speech enhancement in the magnitude~\cite{Tan2018, Ernst2018, Tan2020}, complex time-frequency (T-F)~\cite{Liu2019DeepCASA, Isik2020, Hu2020, Wang2020CSMDereverbJournal, Wang2020aCSMCHiME4, Wang2020css, Taherian2020, Wang2021FCPjournal, Zhao2022FRCRN, Eskimez2022, Taherian2022, Tan2022NSF} and time domain~\cite{Pandey2021}, it usually costs a large amount of computation. %
\item State-of-the-art dual-path models such as DPRNN~\cite{Luo2020} and TF-GridNet~\cite{Wang2022GridNet} are not ideal for real-time enhancement, as they are computationally expensive.
TF-GridNet, for example, runs an LSTM for each frequency at each layer, and at each frame it does not process all the steps in the sequence in parallel. 
\end{itemize}
Equipped with these understandings, we think that recurrent neural networks such as LSTM~\cite{Courville2016} are more suitable for online, low-complexity speech enhancement, because at run time only one past frame needs to be buffered and the memory cost and system complexity can be low.
In addition, small fully-connected blocks (or one-dimensional convolutions) are usually less costly than Conv2D blocks that use large kernels and large input and output channels.

In this context, we investigate using stacked LSTMs as the DNN backbone for frame-online speech enhancement with very low algorithmic latency and complexity.
Although there have been studies exploring this direction~\cite{Zhao2018b, Valin2020, Xia2020, Braun2021, Thakker2022}, they are usually studied in monaural conditions and in teleconferencing scenarios where the allowed processing latency can be as high as $40$ ms~\cite{K.A.Reddy2020} and hence a regularly-large hop size (e.g., 8, 10 and 16 ms) is often used.
In hearing aids setup, however, the requirement on algorithmic latency is usually less than $5$ ms~\cite{ClarityWebpage}.
This means that the hop size cannot go beyond $2.5$ ms if 50\% frame-overlap is used in overlap-add, and such a small hop size would create longer frame sequences to process and requires hardware latency to be less than the hop size in order to realize real-time enhancement. 
In such cases, how to design a low-complexity DNN architecture that can leverage LSTMs to achieve single- and multi-channel enhancement with low algorithmic latency is an important problem to study.

In our experiments, we observe that a multi-layer unidirectional LSTM modeling full-band information performs impressively well even when the hop size is as low as $1$ ms, indicating that LSTM could be very suitable for hearing aids design.
We further integrate the full-band LSTM blocks with sub-band LSTM blocks so that complementary full- and sub-band information can be combined to achieve better enhancement, leading to a novel DNN architecture named FSB-LSTM for STFT-domain speech enhancement with very low algorithmic latency and low complexity.
Evaluation results on single- and multi-channel speech enhancement in noisy-reverberant conditions show the effectiveness of FSB-LSTM over other state-of-the-art low-latency streamable models in the time domain and in the complex T-F domain.
Ablation studies also confirm the effectiveness of the proposed integrated full- and sub-band processing.

\begin{figure}
  \centering  
  \includegraphics[width=8cm]{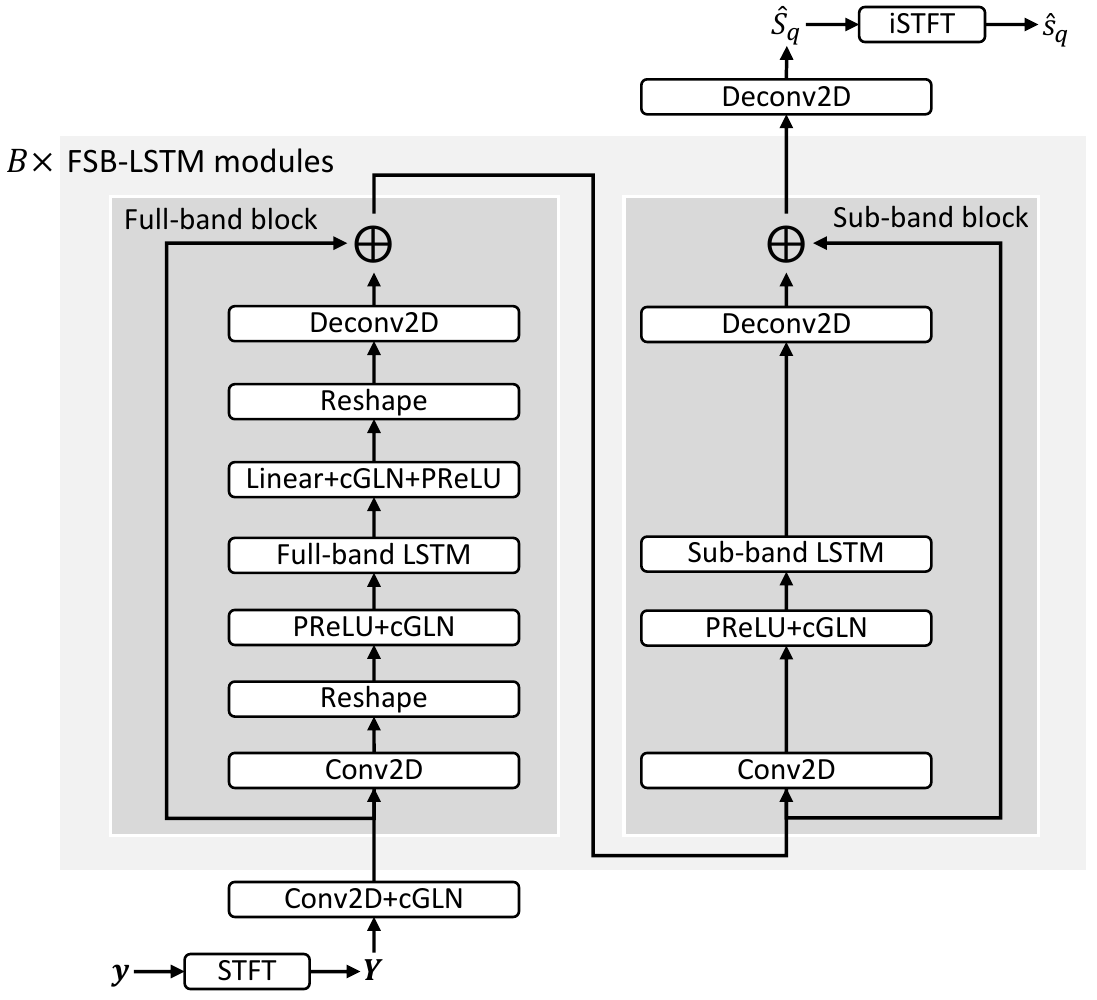}
  \vspace{-0.25cm} \\
  \caption{Overview of proposed system.}
  \label{system_overview_figure}
  \vspace{-0.6cm}
\end{figure}

\vspace{-0.2cm}
\section{Proposed Algorithms}\label{proposeddescription}
\vspace{-0.2cm}

Given a single-speaker, $N$-sample mixture recorded by a $P$-microphone array in noisy-reverberant conditions, the physical model in the time domain can be written as $\mathbf{y}[n] = \mathbf{s}[n] + \mathbf{v}[n]$, where $\mathbf{y}[n]$, $\mathbf{s}[n]$ and $\mathbf{v}[n] \in \RR^P$ respectively denote the mixture, direct-path signal of the target speaker, and non-target signals at sample $n$.
Our study aims at estimating the target direct-path signal captured by a reference microphone $q$ (i.e., $s_q$) based on the mixture in a low-latency, low-complexity setup.

In the STFT domain, we denote the mixture as $\mathbf{Y}(t,f)=\mathbf{S}(t,f)+\mathbf{V}(t,f)\in \CC^P$, where $\mathbf{Y}$, $\mathbf{S}$ and $\mathbf{V}$ are respectively the STFT spectra of $\mathbf{y}$, $\mathbf{s}$ and $\mathbf{v}$, $t$ indexes $T$ frames, and $f$ indexes $F$ frequencies.
Our system operates in the STFT domain.
Following~\cite{Wang2022VeryLowAlgorithmicLatency}, we use a regularly-large input window size (iWS) for STFT and a much smaller output window size (oWS) for overlap-add in inverse STFT (iSTFT), and both iWS and oWS are set to multiples of the hop size (HS).
This way, our system can use an STFT with a regularly-high frequency resolution while still have a low algorithmic latency equal to the smaller oWS rather than the regularly-large iWS.
See~\cite{Wang2022VeryLowAlgorithmicLatency} for the details of this STFT-iSTFT mechanism.

Fig.~\ref{system_overview_figure} illustrates our proposed system.
It is trained to perform multi-microphone complex spectral mapping based speech enhancement~\cite{Wang2020dMCCSMconference, Wang2020css, Tan2022NSF}, where the real and imaginary (RI) components of the mixture $\mathbf{Y}$ are stacked as input features to predict the RI components of target speech $S_q$.
Given an input tensor with shape $2P\times T\times F$, where $2P$ is because we stack the RI components at all the $P$ microphones, we first use a Conv2D layer with kernel size $1\times 3$ along time and frequency to get a $D$-dimensional embedding for each T-F unit, obtaining in a $D\times T\times F$ tensor.
We then use $B$ FSB-LSTM modules, each with a full-band and a sub-band block, to leverage spectral, spatial and temporal information to gradually refine the T-F embeddings.
Next, a 2D deconvolution (Deconv2D) layer with kernel size $1\times 3$ is used to predict the target RI components.
Finally, inverse STFT (iSTFT) is applied for signal re-synthesis.
The loss function is defined on the re-synthesized signal and its magnitude, following the Wav+Mag loss in~\cite{Wang2022VeryLowAlgorithmicLatency}.
The rest of this section describes the full- and sub-band blocks in FSB-LSTM.
To avoid confusion, in Table~\ref{summary_hyperparam} we summarize the hyper-parameters we will use to describe FSB-LSTM.

\begin{table}[t]
\scriptsize
\centering
\sisetup{table-format=2.2,round-mode=places,round-precision=2,table-number-alignment = center,detect-weight=true,detect-inline-weight=math}
\caption{Summary of model hyper-parameters.}
\vspace{-0.25cm}
\label{summary_hyperparam}
\setlength{\tabcolsep}{1.5pt}
\begin{tabular}{
cc
}
\toprule
Symbols & Description \\
\midrule
$B$ & Number of FSB-LSTM modules \\
$D$ & Embedding dimension for each T-F unit \\
\midrule
$E$ & Output channels of Conv2D in full-band blocks \\
$I$ & Kernel size along frequency in Conv2D and Deconv2D in full-band blocks \\
$J$ & Stride size along frequency in Conv2D and Deconv2D in full-band blocks \\
$H$ & Number of hidden units in full-band LSTMs \\
\midrule
$E'$ & Output channels of Conv2D in sub-band blocks \\
$I'$ & Kernel size along frequency in Conv2D and Deconv2D in sub-band blocks \\
$J'$ & Stride size along frequency in Conv2D and Deconv2D in sub-band blocks \\
$H'$ & Number of hidden units in sub-band LSTMs \\
\bottomrule
\end{tabular}
\vspace{-0.6cm}
\end{table}

\vspace{-0.2cm}
\subsection{Full-Band Block}\label{full_band_block}
\vspace{-0.1cm}

Given an input tensor with shape $D\times T\times F$, we compress the $D$-dimensional T-F embeddings within each frame into a frame-level embedding, use an LSTM to refine the frame embedding, and re-compute $D$-dimensional T-F embeddings based on the refined frame embedding.
This way, the LSTM can model all the frequencies at the same time to capture full-band information.

Specifically, we first use a Conv2D layer with input channel $D$, output channel $E$, kernel size $1\times I$, and stride $1\times J$ to compress the $D\times T\times F$ tensor along dimension one and three to $E\times T\times (\frac{Q-I}{J}+1)$, after zero-padding the frequency dimension to $Q= \lceil \frac{F-I}{J} \rceil \times J + I$.
We then reshape it to a 2D tensor by flattening the first and third dimensions to obtain a tensor with shape $T\times A$ with $A=E\times(\frac{Q-I}{J}+1)$, apply PReLU, and perform causal global layer normalization (cGLN)~\cite{Luo2019}, which computes the mean and variance for normalization based on both dimensions in a causal way and uses two $A$-dimensional vectors to respectively scale and shift along the first dimension.
Next, we use an LSTM with $H$ hidden units to model the $A$-dimensional frame embeddings, obtaining a tensor with shape $T\times H$.
After that, a linear layer is applied to map the $H$-dimensional embedding to $A$-dimensional, followed by cGLN and PReLU.
Finally, we reshape the $T\times A$ tensor back to $E\times T\times (\frac{Q-I}{J}+1)$, and use a Deconv2D layer with input channel $E$, output channel $D$, kernel size $1\times I$, and stride $1\times J$ to compute a $D\times T\times F$ tensor, which is added to the original input tensor to this full-band block via a residual connection, after removing padded zeros.

\vspace{-0.2cm}
\subsection{Sub-Band Block}\label{sub_band_block}
\vspace{-0.1cm}

In~\cite{Wang2022GridNet, Wang2022GridNetJournal}, we find that using sub-band modules to leverage sub-band information is very effective at dereverberation and leveraging spatial information.
However, TF-GridNet proposed in~\cite{Wang2022GridNet, Wang2022GridNetJournal} runs a sub-band module at each frequency, consuming a large amount of computation.
To reduce the computation, we reduce the number of frequencies by using convolution based down-sampling, and use much fewer input, hidden and output units in LSTMs.

In detail, given an input tensor with shape $D\times T\times F$, we first use a Conv2D layer with input channel $D$, output channel $E'$, kernel size $1\times I'$, and stride size $1\times J'$ to down-sample the $D\times T\times F$ tensor along dimension one and three to $E'\times T\times (\frac{Q'-I'}{J'}+1)$, after zero-padding the frequency dimension to $Q'= \lceil \frac{F-I'}{J'} \rceil \times J' + I'$.
Next, we apply PReLU, and perform cGLN, which, in the case of 3D tensors, computes the mean and variance for normalization based on all the three dimensions in a causal way and uses two $E'$-dimensional vectors to respectively scale and shift along the first dimension.
After that, we view the tensor as $\frac{Q'-I'}{J'}+1$ sequences, each with length $T$, and use an LSTM with $H'$ hidden units to refine the $E'$-dimensional T-F embeddings, obtaining a tensor with shape $H'\times T\times (\frac{Q'-I'}{J'}+1)$.
Note that the LSTM is shared across all the sub-bands to reduce model parameters.
At last, we use a Deconv2D layer with input channel $H'$, output channel $D$, kernel size $1\times I'$, and stride size $1\times J'$ to compute a $D\times T\times F$ tensor, which is added to the input tensor of the sub-band block through a residual connection, after removing padded zeros.

\vspace{-0.2cm}
\subsection{Discussion on Network Design} \label{discussion_net_design}
\vspace{-0.1cm}

In our network, we maintain an \textit{information highway}, which flows an over-complete T-F representation (i.e., the $D$-dimensional T-F embeddings) of multi-channel input signals inside the network through residual connections.
When refining the T-F embeddings, we first perform down-sampling to extract desired discriminative features, then use LSTM layers to perform full- or sub-band temporal modeling, and finally up-sample and add it back to the input tensor.
This could be a good strategy especially for multi-microphone enhancement, as it can maintain the fine-grained information of multiple input signals (e.g., spectro-temporal and spatial patterns) and at the same time extract different discriminative information of interest at different blocks for better enhancement.

In comparison, an early popular way of using stacked LSTMs feeds $2P\times F$-dimensional input features directly to a multi-layer LSTM (in a way similar to that in the deep clustering~\cite{Hershey2016, Wang2018aMCDC} and permutation invariant training~\cite{Kolbak2017} studies).
However, since the hidden dimension of LSTM is usually much smaller than the input dimension when $P$ is large, the resulting model would be limited at exploiting spatial information due to the compression of input features.
Similarly, modern time-domain models such as Conv-TasNet~\cite{Luo2019, ZhangJisi2020} tend to create a bottleneck representation immediately after the encoder.
Such a bottleneck could lead to loss of information when the input dimension is high (e.g., in multi-microphone cases).
One solution is to use recurrent U-Net based models for multi-channel separation~\cite{Wang2020dMCCSMconference, Wang2020css, Tan2022NSF}, where the lower layers in the U-Net encoder (and the corresponding layers in the decoder) can have an over-complete representation of input features to maintain fine-grained patterns, and the input features are gradually down-sampled to a dimension suitable for recurrent networks.
In our experiments, we will show that FSB-LSTM produces better performance than a strong recurrent U-Net based model~\cite{Wang2021LowDistortion} and a multi-channel Conv-TasNet~\cite{ZhangJisi2020, Tu2021}.

All the convolutions in our models have a kernel size of one along time.
This way, we avoid buffering past frames due to the use of causal convolution, and just use LSTMs to model temporal information.
The Conv2D and Deconv2D layers in our models are not very costly, as they are used with a large stride size, a small kernel size, and few input and output channels.
For deconvolution, we use a custom implementation\footnote{
Deconvolution (\textit{a.k.a} transposed convolution)~\cite{Courville2016} is typically implemented by first interleaving zeros to the input tensor based on the stride size and then performing regular convolution.
This increases the MAC operations when the stride is larger than one, because the new input tensor would have more elements to convolve due to the interleaved zeros.
In our study, we implement deconvolution as a linear layer followed by overlap-add along frequency (please do not confuse this overlap-add with that in iSTFT).
This way, we can save the computation wasted on the interleaved zeros. On the other hand, the overlap-add usually costs negligible MAC operations compared to the linear layer.
The number of MAC operations of our implementation is roughly $1/J$ of that of the typical implementation, where $J$ is the stride size.
} to reduce the number of multiply–accumulate (MAC) operations.

We emphasize that the proposed network only needs to buffer the hidden and cell states of LSTMs in the past frame.
The run-time memory cost and complexity of maintaining the buffer is low.

\vspace{-0.2cm}
\section{Experimental Setup}\label{setup}
\vspace{-0.2cm}

We validate our algorithms on a simulated noisy-reverberant speech enhancement task.
This section describes the dataset, system configurations, baseline systems, and evaluation metrics.

\vspace{-0.2cm}
\subsection{Dataset}
\vspace{-0.1cm}

We use a simulated data, which was used in recent studies~\cite{Wang2021LowDistortion, Wang2022VeryLowAlgorithmicLatency}, to evaluate the proposed algorithms.
Using the split of clean speech in WSJCAM0, the dataset simulates 39,245 ($\sim$77.7 h), 2,965 ($\sim$5.6 h) and 3,260 ($\sim$8.5 h) noisy-reverberant mixtures respectively for training, validation and testing.
The clips in the development set of FSD50k~\cite{Fonseca2020} are sampled to simulate the noises for training and validation, and those in the evaluation set for testing.
Each simulated mixture contains up to seven noise clips, with one longer than ten seconds as background and the others as foreground noises.
The simulated microphone array contains six microphones arranged uniformly on a circle with a diameter of 20 cm.
The direction of each source to the array center is sampled from the range $[0,2\pi)$, distance from $[0.75, 2.5]$ m, and the reverberation time from $[0.2, 1.0]$ s.
We treat each sound source as a point source, convolve each source with a simulated room impulse response, and summate the convolved sources to create the mixture.
The signal-to-noise ratio between the target direct-path speech and reverberant noise is drawn from $[-8, 3]$ dB.
The sampling rate is 16 kHz.
For two-channel processing, we use signals at the first and the fourth microphones; and for monaural processing, the first microphone is used.
The target direct-path signal captured at the first microphone is used as the label for model training and as the reference for metric computation.

\vspace{-0.2cm}
\subsection{System Configurations}
\vspace{-0.1cm}

We aim at an enhancement system with an algorithmic latency of $4$ ms, which is slightly shorter than the 5 ms requirement suggested in the recent Clarity challenge~\cite{ClarityWebpage} proposed for hearing aids design.
For STFT and iSTFT, in default the iWS is set to $16$ ms, HS to $2$ ms, and oWS to $4$ ms, resulting in an algorithmic latency of $4$ ms~\cite{Wang2022VeryLowAlgorithmicLatency}.
The rectangular window is used as the analysis window.
Given a sampling rate of 16 kHz, a $256$-point discrete Fourier transform is used to extract $129$-dimensional complex spectra at each frame.
Through the validation set, we set $B=3$, $D=32$, $E=8$, $I=8$, $J=4$, $H=256$ $E'=64$, $I'=5$, $J'=5$, and $H'=64$ (see Table~\ref{summary_hyperparam} for the definition of the notations).
In this configuration, there are $26$ sub-bands and the MAC operations of the sub-band module are around twice as many as the full-band module.

\vspace{-0.2cm}
\subsection{Baseline Systems}
\vspace{-0.1cm}

We consider Conv-TasNet~\cite{Luo2019}, its multi-channel extension MC-Conv-TasNet~\cite{ZhangJisi2020, Tu2021}, LSTM-ResUNet~\cite{Wang2022VeryLowAlgorithmicLatency}, and a full-band only LSTM model as the major baselines.
All of them are trained with the same loss as FSB-LSTM.

Conv-TasNet~\cite{Luo2019} is an excellent time-domain model in speech separation. 
It uses learned bases on very short windows of signals to achieve separation with very low algorithmic latency.
Using the symbols listed in Table I of the Conv-TasNet paper~\cite{Luo2019}, we set the hyper-parameters of Conv-TasNet and MC-Conv-TasNet to $N=512, B=158, S_c=158, H=512, P=3, X=8$, and $R=3$ (please do not confuse these symbols with those defined in this paper).
$B$ and $S_c$ are set slightly larger than the default $128$ suggested in~\cite{Luo2019}, considering that in MC-Conv-TasNet there are additional spatial embeddings concatenated to the spectral embeddings as the input to the separator of Conv-TasNet.
Following~\cite{ZhangJisi2020, Tu2021}, the spatial embedding dimension is set to $60$ for two-channel enhancement and to $360$ for six-channel enhancement.

LSTM-ResUNet~\cite{Wang2022VeryLowAlgorithmicLatency} is a representative complex T-F domain model, consisting of a multi-layer LSTM sandwiched by a UNet with residual net blocks inserted at multiple frequency scales.
It uses a shorter oWS than the iWS in overlap-add to realize enhancement with low algorithmic latency~\cite{Wang2022VeryLowAlgorithmicLatency}.
We emphasize that its network architecture shares many similarities with recent complex T-F domain models~\cite{Tan2020, Hu2020, Liu2019DeepCASA, Tan2022NSF} in speech enhancement.
We therefore consider it as a major baseline in addition to Conv-TasNet.

To show the effectiveness of including sub-band LSTMs, we replace the sub-band module in Fig.~\ref{system_overview_figure} with the full-band module.
This way, the system essentially stacks multiple full-band LSTMs.
We denote this system as \textbf{FB-LSTM}, where ``FB'' means full-band.
We experiment FB-LSTM with $6$ and $9$ full-band LSTM blocks, since we use $B=3$ full- and sub-band modules in FSB-LSTM (totalling $2\times 3=6$ LSTMs) and each sub-band module costs roughly twice as many MAC operations as the full-band module.

\vspace{-0.2cm}
\subsection{Evaluation Metrics}
\vspace{-0.1cm}

The evaluation metrics include scale-invariant signal-to-distortion ratio (SI-SDR), perceptual evaluation of speech quality (PESQ), and extended short-time objective intelligibility (eSTOI).
For PESQ, we use the \textit{python-pesq} (v0.0.2) toolkit to report narrow-band MOS-LQO scores.
The number of model parameters is reported in millions (M).
Using the \textit{ptflops} toolkit, we report the amount of computation by counting MAC in giga-operations per second (GMAC/s).

\vspace{-0.2cm}
\section{Evaluation Results}\label{results}

\begin{table}
\scriptsize
\centering
\sisetup{table-format=2.2,round-mode=places,round-precision=2,table-number-alignment = center,detect-weight=true,detect-inline-weight=math}
\caption{Results of FB-LSTM at various hop sizes (6ch).}
\vspace{-0.25cm}
\label{LSTM_with_hop_6ch_results}
\setlength{\tabcolsep}{2pt}
\begin{tabular}{
l
S[table-format=2,round-precision=0]
S[table-format=2,round-precision=0]
S[table-format=2,round-precision=0]
S[table-format=1.2,round-precision=2]
S[table-format=1.2,round-precision=2]
S[table-format=3.1,round-precision=1]
S[table-format=1.2,round-precision=2]
S[table-format=1.3,round-precision=3]
}
\toprule
{\multirow{2}{*}{Systems}} & {iWS} & {oWS} & {HS} & {\#params} & {\multirow{2}{*}{GMAC/s}} & {SI-SDR} & {\multirow{2}{*}{PESQ}} & {\multirow{2}{*}{eSTOI}} \\
 & {(ms)} & {(ms)} & {(ms)} & {(M)} & & {(dB)} & & \\
\midrule
Unprocessed & {-} & {-} & {-} & {-} & {-} & -6.2 & 1.44 & 0.411 \\
\midrule
FB-LSTM (6-layer) & 16 & 16 & 8 & 3.59 & 0.58203 & 3.87274 & 2.057583 & 0.721020 \\
FB-LSTM (6-layer) & 16 & 8 & 4 & 3.59 & 1.1640420 & 5.766651 & 2.2831158156409583 & 0.7755472049 \\
FB-LSTM (6-layer) & 16 & 4 & 2 & 3.59 & 2.325525 & 6.78903 & 2.36796352 & 0.7948465391426 \\
FB-LSTM (6-layer) & 16 & 2 & 1 & 3.59 & 4.6485 & \bfseries 7.961082959959021 & \bfseries 2.535298190453301 & \bfseries 0.8209161871480822 \\ %
\bottomrule
\end{tabular}
\vspace{-0.25cm}
\end{table}

\begin{table}
\scriptsize
\centering
\sisetup{table-format=2.2,round-mode=places,round-precision=2,table-number-alignment = center,detect-weight=true,detect-inline-weight=math}
\caption{Results on speech enhancement (6ch).}
\vspace{-0.25cm}
\label{6ch_results}
\setlength{\tabcolsep}{2.5pt}
\begin{tabular}{
l
S[table-format=1.2,round-precision=2]
S[table-format=1.2,round-precision=2]
S[table-format=3.1,round-precision=1]
S[table-format=1.2,round-precision=2]
S[table-format=1.3,round-precision=3]
}
\toprule
Systems & {\#params (M)} & {GMAC/s} & {SI-SDR (dB)} & {PESQ} & {eSTOI} \\
\midrule
Unprocessed & {-} & {-} & -6.2 & 1.44 & 0.411 \\
\midrule
FSB-LSTM & 1.96 & 3.373 & \bfseries 7.79101 & \bfseries 2.61094015 & \bfseries 0.8299041 \\
FB-LSTM (6-layer) & 3.59 & 2.32552 & 6.78903 & 2.36796352 & 0.7948465391426 \\
FB-LSTM (9-layer) & 5.38 & 3.428025 & 7.5873158274588395 & 2.5081466446624945 & 0.8162881265313816 \\
\midrule
MC-Conv-TasNet~\cite{ZhangJisi2020, Tu2021} & 6.37 & 3.7625 & 5.2 & 2.24 & 0.764 \\
LSTM-ResUNet~\cite{Wang2022VeryLowAlgorithmicLatency} & 2.33 & 3.539571 & 6.19791 & 2.2300008 & 0.767019 \\
\bottomrule
\end{tabular}
\vspace{-0.5cm}
\end{table}

\vspace{-0.2cm}
\subsection{Effectiveness of LSTM at Dealing with Small Hop Sizes}\label{results_lstm_small_hop}
\vspace{-0.1cm}

Although LSTM has been criticized for not being good enough at modeling long sequences resulted from small hop sizes~\cite{Luo2019}, in our experiments (which focus on frame-online enhancement) we find FB-LSTM performing surprisingly well even if the hop size is as low as 1 ms.
See Table~\ref{LSTM_with_hop_6ch_results} for the results.
The iWS is always $16$ ms.
We reduce HS together with oWS so that the frame-overlap ratio in overlap-add is always 50\% when the algorithmic latency (equal to oWS) becomes smaller.
Every time HS is halved, the amount of computation is approximately doubled as the number of frames to process is doubled.
From the results, we observe that a smaller HS (and oWS) leads to better performance, even though the resulting frame sequence gets much longer and the future context information that can be utilized (up to oWS to the future) becomes less.

Although using smaller hop sizes was found effective in time-domain speaker separation studies such as Conv-TasNet~\cite{Luo2019} and DPRNN~\cite{Luo2020}, they are often used with more advanced architectures rather than simple uni-directional LSTMs that model frame sequences from left to right.
Our study observes that such simple causal LSTMs can perform reasonably well for hop sizes as low as 1 ms.
This finding is very significant, as it indicates that simple LSTMs, which have very low run-time complexity, can produce promising enhancement results in a hearing aid setup which requires very low processing latency.

\begin{table}
\scriptsize
\centering
\sisetup{table-format=2.2,round-mode=places,round-precision=2,table-number-alignment = center,detect-weight=true,detect-inline-weight=math}
\caption{Results on speech enhancement (2ch).}
\vspace{-0.25cm}
\label{2ch_results}
\setlength{\tabcolsep}{2.5pt}
\begin{tabular}{
l
S[table-format=1.2,round-precision=2]
S[table-format=1.2,round-precision=2]
S[table-format=3.1,round-precision=1]
S[table-format=1.2,round-precision=2]
S[table-format=1.3,round-precision=3]
}
\toprule
Systems & {\#params (M)} & {GMAC/s} & {SI-SDR (dB)} & {PESQ} & {eSTOI} \\
\midrule
Unprocessed & {-} & {-} & -6.2 & 1.44 & 0.411 \\
\midrule
FSB-LSTM & 1.96 & 3.31277 & \bfseries 4.91881 & \bfseries 2.19990 & \bfseries 0.7526811 \\
FB-LSTM (6-layer) & 3.59 & 2.26527 & 4.24713340 & 2.074001 & 0.72877927 \\
FB-LSTM (9-layer) & 5.38 & 3.36777 & 4.451304537600886 & 2.142454030652719 & 0.7441581478162531 \\
\midrule
MC-Conv-TasNet~\cite{ZhangJisi2020, Tu2021} & 6.19 & 3.67764 & 3.56102 & 1.9950 & 0.711390 \\
LSTM-ResUNet~\cite{Wang2022VeryLowAlgorithmicLatency} & 2.33 & 3.4828 & 4.25184 & 2.0581 & 0.72587 \\
\bottomrule
\end{tabular}
\vspace{-0.25cm}
\end{table}

\begin{table}
\scriptsize
\centering
\sisetup{table-format=2.2,round-mode=places,round-precision=2,table-number-alignment = center,detect-weight=true,detect-inline-weight=math}
\caption{Results on speech enhancement (1ch).}
\vspace{-0.25cm}
\label{1ch_results}
\setlength{\tabcolsep}{3.5pt}
\begin{tabular}{
l
S[table-format=1.2,round-precision=2]
S[table-format=1.2,round-precision=2]
S[table-format=3.1,round-precision=1]
S[table-format=1.2,round-precision=2]
S[table-format=1.3,round-precision=3]
}
\toprule
Systems & {\#params (M)} & {GMAC/s} & {SI-SDR (dB)} & {PESQ} & {eSTOI} \\
\midrule
Unprocessed & {-} & {-} & -6.2 & 1.44 & 0.411 \\
\midrule
FSB-LSTM & 1.96 & 3.29771 & \bfseries 3.0842705 & \bfseries 1.9161919 & \bfseries 0.6879391473 \\
FB-LSTM (6-layer) & 3.59 & 2.2502 & 2.553 & 1.843008 & 0.670816 \\
FB-LSTM (9-layer) & 5.38 & 3.3527 & 2.45907 & 1.834456580 & 0.666514 \\
\midrule
Conv-TasNet~\cite{Luo2019} & 6.18 & 3.668857 & 2.1584 & 1.7841 & 0.65729 \\
LSTM-ResUNet~\cite{Wang2022VeryLowAlgorithmicLatency} & 2.32 & 3.469585 & 2.8341 & 1.8997 & 0.68235 \\
\bottomrule
\end{tabular}
\vspace{-0.5cm}
\end{table}

\vspace{-0.2cm}
\subsection{Results of FSB-LSTM}\label{results_proposed}
\vspace{-0.1cm}

Table~\ref{6ch_results}, \ref{2ch_results} and \ref{1ch_results} respectively present the results of FSB-LSTM on six-, two- and one-channel speech enhancement.
We can see that, with a smaller model size and using fewer MAC/s operations, FSB-LSTM produces better enhancement than Conv-TasNet, MC-Conv-TasNet and LSTM-ResUNet.
Note that both Conv-TasNet and LSTM-ResUNet contain convolutions dilated along time and need to buffer many past frames at run time. %

FSB-LSTM produces better results than 6-layer and 9-layer FB-LSTM. This shows the benefits of using the sub-band blocks.

\vspace{-0.2cm}
\subsection{Run-Time Complexity}\label{buffer_size}
\vspace{-0.1cm}

We compute the run-time buffer size of each model in an online, streaming setup, based on single-precision floating-point operations.
FSB-LSTM only needs to buffer LSTMs' hidden and cell states in the past frame.
It has a buffer size of $46.2$ kilobytes (KB) to maintain, while the buffer sizes of Conv-TasNet and LSTM-ResUNet are respectively $3133.4$ and $1815.1$ KB.
Such a small buffer size makes it possible to have the buffered tensors reside in a higher cache hierarchy, which has very limited space (e.g., tens of KB at Level 1 and several MB at Level 2) even in modern processors.
The small buffer size and the low algorithmic complexity also make it easier for the hardware latency to be smaller than the hop size to realize real-time enhancement in resource-constrained hearing-aid scenarios.

\vspace{-0.2cm}
\section{Conclusion}\label{conclusion}
\vspace{-0.2cm}

We have proposed a novel FSB-LSTM architecture that integrates full- and sub-band modeling for low-complexity, low-algorithmic-latency speech enhancement.
Our experiments show that FSB-LSTM outperforms previously proposed state-of-the art streamable, low-latency models with much less buffer memory and less computational burden in MAC operations. 
Future research will further reduce algorithmic latency and explore DNN quantization and distillation to further reduce complexity, at the same time maintaining a strong enhancement performance.

\bibliographystyle{IEEEtran}
{\footnotesize
\bibliography{references.bib}
}

\end{document}